\documentstyle[11pt,epsfig]{article}

\catcode`\^^?=9 

\def\be{\begin{equation}}
\def\ee{\end{equation}}
\def\ba{\begin{eqnarray}}
\def\ea{\end{eqnarray}}
\def\mref#1{Eq.~(\ref{eq:#1})}

\def\oref#1{\ref{eq:#1}}
\def\mlab#1{\label{eq:#1}}

\def\slash#1{/\llap #1}

\def\G{{\cal G}}
\def\S{\Sigma}
\def\F{{\cal F}}
\def\Tr{\mbox{Tr}}
\def\P{{\cal P}}

\def\im{i}

\begin{document}

\noindent December, 1994 \hfill DTP-94/100

\vskip2cm

\begin{center}

{\huge Numerical cancellation of photon quadratic divergence in the
study of the Schwinger-Dyson equations in Strong Coupling QED}

\vskip1cm

{\large \bf J.C.R.~Bloch and M.R.~Pennington}\\
{Centre for Particle Theory,\\
University of Durham,\\
Durham DH1 3LE, U.K.}

\vskip1cm

{\parbox{12cm}{{\bf Abstract~:} \noindent \sf The behaviour of the
photon renormalization function in strong coupling QED has been
recently studied by Kondo, Mino and Nakatani.  We find that the sharp
decrease in its behaviour at intermediate photon momenta is an
artefact of the method used to remove the quadratic divergence in the
vacuum polarization. We discuss how this can be avoided in numerical
studies of the Schwinger-Dyson equations.  }}

\end{center}

\pagestyle{empty}

\newpage

\baselineskip=5mm

\pagestyle{plain}

As part of a longer study of chiral symmetry breaking in strong QED
with $N_f$ flavours we have turned our attention to the results of
\cite{Kondo92} where a solution of the simultaneous Schwinger-Dyson
equations in strong coupling QED is presented in a self-consistent
way. Previous studies have most commonly approximated the photon
propagator by its one loop perturbative form in undertaking either
analytic \cite{Kondo91,Gusynin} or numerical calculations
\cite{Kondo92b}. In contrast, Kondo et al. \cite{Kondo92} have studied
a fully coupled system of equations for the photon and fermion
propagators.  Then the photon renormalization function is determined in
a way that is claimed to be self-consistent.

\begin{figure}[h]
\begin{center}
{}~\epsfig{file=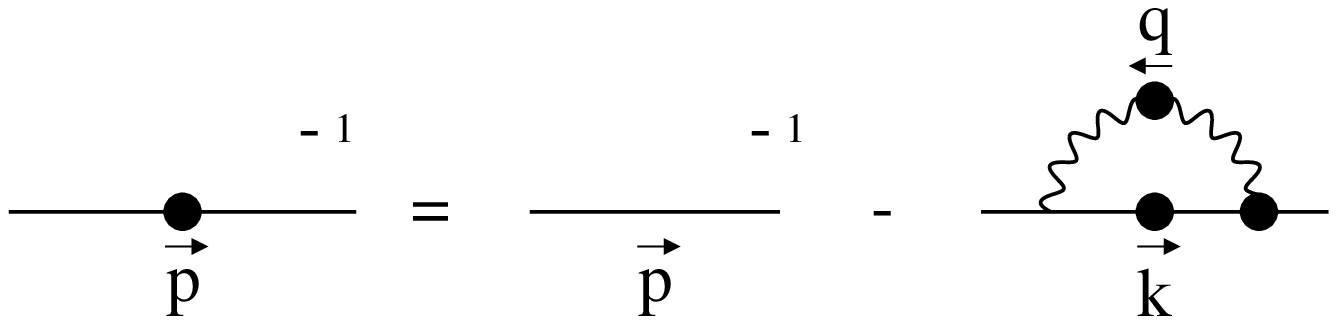,width=15cm}
{}~\epsfig{file=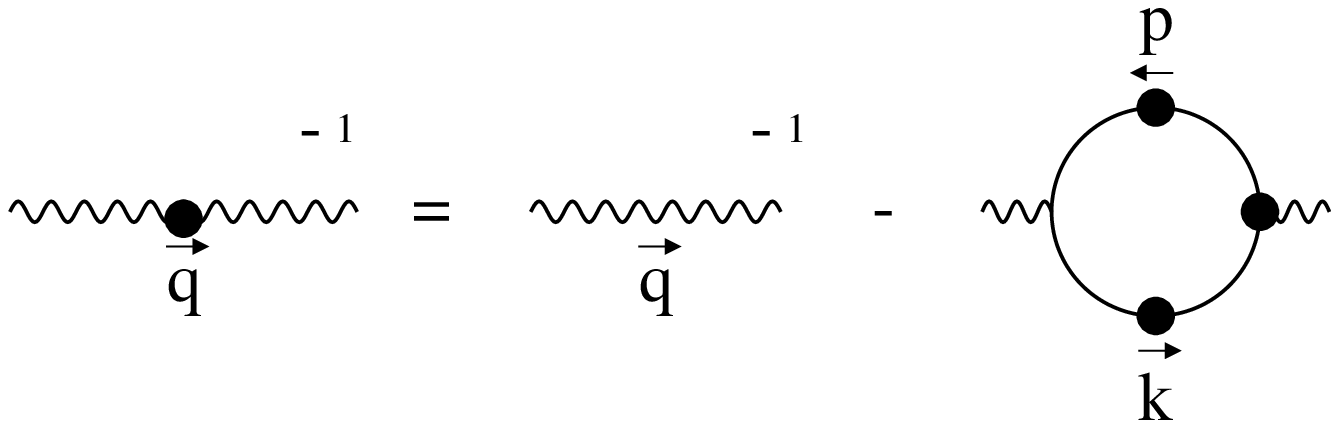,width=15cm}
\end{center}
\vspace{-1cm}
\caption{Schwinger-Dyson equations for the fermion and photon propagator.}
\label{fig:figSD}
\end{figure}

The Schwinger-Dyson equations for the fermion propagator and for the
photon propagator in QED are given diagrammatically in Fig.~\ref{fig:figSD}.

\vspace{0.5cm}
Substituting $\im S_F$ for the fermion propagator, $\im D_{\mu\nu}$
for the photon propagator and $(-\im e \Gamma^\mu)$ for the vertex
yields~:
\be
\Big[\im S_F(p)\Big]^{-1} = \Big[\im S_F^0(p)\Big]^{-1} -
\frac{e^2}{(2\pi)^4} \int d^4k \,
(\im\Gamma^\mu(k,p)) \im S_F(k) (\im\gamma^\nu) \im D_{\mu\nu}(q) \, ,
\mlab{fermion}
\ee
where $q = k-p$, and
\be
\Big[\im D_{\mu\nu}(q)\Big]^{-1} = \Big[\im D_{\mu\nu}^0(q)\Big]^{-1}
- \frac{(-1)N_f e^2}{(2\pi)^4}
\int d^4k \, \Tr \left[ (\im\Gamma_\mu(k,p)) \im S_F(k)
(\im\gamma_\nu) \im S_F(p) \right] \, ,
\mlab{photon}
\ee
where $p = k-q$.

We define the full fermion propagator of momentum $p$ by~~:
\[
\im S_F(p) = \frac{\im \F(p^2)}{\slash{p} - \S(p^2)} \, .
\]
The bare fermion propagator is~:
\[
\im S_F^0(p) = \frac{\im}{\slash{p} - m_0} \, .
\]

The full photon propagator of momentum $q$ is given by~:
\be
\im D_{\mu\nu}(q) = -\frac{\im}{q^2} \left[ \G(q^2)\left(g_{\mu\nu}-\frac{q_\mu
q_\nu}{q^2}\right) + \xi \frac{q_\mu q_\nu}{q^2}\right] \, .
\mlab{fullD}
\ee
The bare photon propagator is~:
\be
\im D_{\mu\nu}(q) = -\frac{\im}{q^2} \left[\left(g_{\mu\nu}-\frac{q_\mu
q_\nu}{q^2}\right) + \xi \frac{q_\mu q_\nu}{q^2}\right] \, .
\mlab{bareD}
\ee

 From \mref{fermion} one can project out the integral equations for
$\S(p^2)$ and $\F(p^2)$. In Minkowski space these are given by~:
\ba
\frac{\S(p^2)}{\F(p^2)} &=& m_0 - \frac{\im e^2}{4(2\pi)^4} \int d^4k \,
\frac{\F(k^2)}{(k^2-\S^2(k^2))q^2} \\
&&\times \left[ \G(q^2)\left(g_{\mu\nu}-\frac{q_\mu q_\nu}{q^2}\right)
 + \xi \frac{q_\mu q_\nu}{q^2} \right]
\Tr[\Gamma^\mu(k,p)(\slash{k}+\S(k^2)) \gamma^\nu] \, ,  \nonumber
\ea
\ba
\frac{1}{\F(p^2)} &=& 1 + \frac{\im e^2}{4(2\pi)^4} \frac{1}{p^2} \int d^4k \,
\frac{\F(k^2)}{(k^2-\S^2(k^2))q^2} \\
&&\times \left[ \G(q^2)\left(g_{\mu\nu}-\frac{q_\mu q_\nu}{q^2}\right)
 + \xi \frac{q_\mu q_\nu}{q^2} \right]
\Tr[\slash{p}\Gamma^\mu(k,p)(\slash{k}+\S(k^2)) \gamma^\nu] \, . \nonumber
\ea

It is important to note that unless the vertex $\Gamma^{\mu}(k,p)$
satisfies the Ward-Takahashi identity and the regularization of the
loop integrals is translation invariant, the photon propagator of
\mref{photon} will {\bf not} have the Lorentz structure of \mref{fullD}
with the coefficients of $g^{\mu\nu}$ and $q^{\mu}q^{\nu}$ being
related to a single function $\G(q^2)$. When these conditions are
satisfied then the integral equation for $\G(q^2)$ can be deduced by
applying the projection operator $\P_{\mu\nu} = g_{\mu\nu}- n {q_\mu
q_\nu}/{q^2}$ (with any value of $n$) to \mref{photon}~:
\ba
\frac{1}{\G(q^2)} &=& 1 - \frac{\im N_f e^2}{3(2\pi)^4}\frac{1}{q^2}
\int d^4k \, \frac{1}{(k^2-\S^2(k^2))(p^2-\S^2(p^2))} \mlab{G1} \\
&&\hspace{3cm} \times \, \P_{\mu\nu} \Tr[\Gamma^\mu(k,p)(\slash{k}+\S(k^2))
\gamma^\nu(\slash{p}+\Sigma(p^2))] \,. \nonumber
\ea

In general, if we regularize the theory using an ultraviolet cutoff
the vacuum polarization integral in \mref{G1} contains a quadratic
divergence which has to be removed, since such a photon mass term is
not allowed in more than 2 dimensions. One can show that the $q_\mu
q_\nu/q^2$ term of the transverse part cannot receive any quadratic
divergent contribution. Consequently, if we choose the projection
operator $\P_{\mu\nu}$ of \mref{G1} with $n=4$, the resulting integral
will be free of quadratic divergences because the contraction
$\P_{\mu\nu} g^{\mu\nu}$ vanishes.

A much used alternative procedure is to take the projection operator
in its simplest form, $\P_{\mu\nu} = g_{\mu\nu}$. The resulting vacuum
polarization integral then contains a quadratic divergence which can
be removed explicitly by imposing~:
\be
\lim_{q^2 \rightarrow 0} \;\frac{q^2}{\G(q^2)} = 0 \, ,
\mlab{limit}
\ee
to ensure a massless photon. If we write the photon renormalization
function as~:
\[
\G(q^2) = \frac{1}{1 + \Pi(q^2)} \, ,
\]
\mref{limit} then corresponds to a renormalization of the vacuum polarization
$\Pi(q^2)$~:
\be
q^2 \tilde{\Pi}(q^2) = q^2 \Pi(q^2) - \lim_{q^2 \rightarrow 0} q^2
\Pi(q^2) \, .
\mlab{vacpolrenorm}
\ee

This is the procedure adopted by Kondo et al. \cite{Kondo92}. They
solve numerically the coupled set of integral equations for the
dynamical fermion mass $\S(p^2)$ and the photon renormalization
function $\G(q^2)$ in the case of zero bare mass, $m_0 \equiv 0$. The
calculations are performed in the Landau gauge with the bare vertex
approximation, i.e.  $\Gamma^\mu(k,p)
\equiv \gamma^\mu$. As a further approximation they decouple the
$\F$-equation by putting $\F(p^2) \equiv 1$. While the quadratic divergence
in the vacuum polarization is removed by imposing \mref{vacpolrenorm},
the fact that the Ward-Takahashi identity is not satisfied, when
dynamical mass is generated, makes the results procedure dependent.

The integral equations one obtains using these approximations,
transformed to Euclidean space, changing to spherical coordinates and
introducing an ultraviolet cutoff $\Lambda^2$ on the radial
integrals, are given by~:
\be
\S(p^2) = \frac{3\alpha}{2\pi^2} \int_0^{\Lambda^2} dk^2
\frac{k^2\S(k^2)}{k^2 + \S^2(k^2)} \int_0^\pi d\theta \sin^2\theta
\frac{\G(q^2)}{q^2} \, ,
\mlab{S2}
\ee
where $q^2 = p^2 + k^2 - 2pk\cos\theta$, with $p=\sqrt{p^2}$,
$k=\sqrt{k^2}$, and
\vspace{2mm}
\ba
\lefteqn{\frac{1}{\G(q^2)} = 1 - \frac{4N_f\alpha}{3\pi^2}\frac{1}{q^2}
\int_0^{\Lambda^2} dk^2 \frac{k^2}{k^2+\S^2(k^2)}
\int_0^\pi d\theta \sin^2\theta} \mlab{G2} \\
&&\times \left\{ \frac{k^2 - kq\cos\theta + 2\S(k^2)\S(p^2)}{p^2+\S^2(p^2)} -
\frac{k^2+2\S^2(k^2)}{k^2+\S^2(k^2)} \right\} \, ,\nonumber
\ea
where $p^2 = q^2 + k^2 - 2qk\cos\theta$, with $q=\sqrt{q^2}$, $k=\sqrt{k^2}$.

The second term in $\{\cdots\}$ in \mref{G2} subtracts the quadratic
divergence. Recall that in QED the momentum dependence of the coupling
comes wholly from the photon renormalization function, so solutions
for $\G(q^2)$ give the running of the coupling. Kondo et al. solve this
coupled set of non-linear integral equations,
Eqs.~(\oref{S2}, \oref{G2}), for $N_f=1$ and find a symmetry breaking
phase for $\alpha$ greater than some critical coupling $\alpha_c
\approx 2.084$.

\begin{figure}
\begin{center}
{}~\epsfig{file=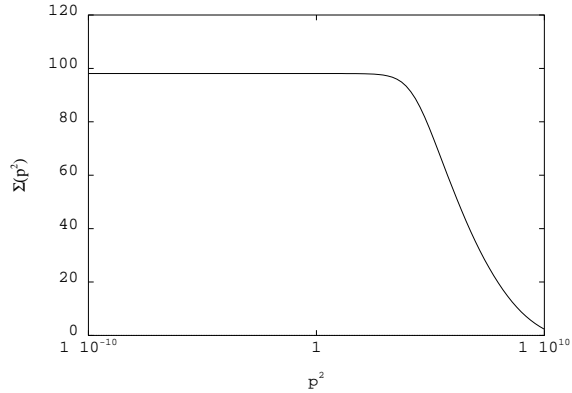,height=8cm,angle=-90}
\end{center}
\vspace{-0.5cm}
\caption{Dynamical mass function $\S(p^2)$, as a function of momentum
$p^2$ for $N_f=1$ and $\alpha=2.086$ as calculated in a
self-consistent way as in \protect{\cite{Kondo92}} ($\Lambda = 10^5$).}
\label{fig:figB}
\end{figure}

\begin{figure}
\begin{center}
{}~\epsfig{file=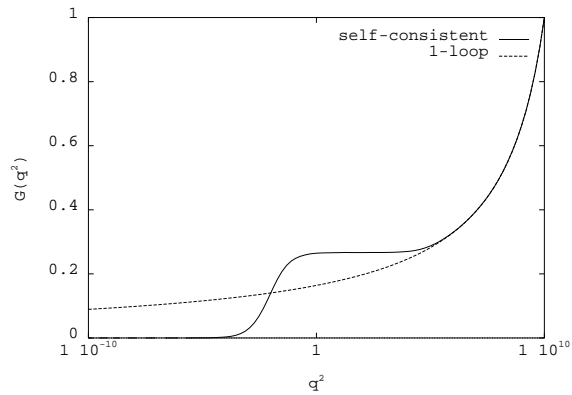,height=8cm,angle=-90}
\end{center}
\vspace{-0.5cm}
\caption{Photon renormalization function $\G(q^2)$, as a function of momentum
$q^2$ for $N_f=1$ and $\alpha=2.086$ as calculated in a
self-consistent way as in \protect{\cite{Kondo92}} and in 1-loop
approximation ($\Lambda = 10^5$).}
\label{fig:figG}
\end{figure}

In Figs.~\ref{fig:figB}, \ref{fig:figG} We display the results for a
value of $\alpha = 2.086$, close to its critical value. The dynamical
mass function, $\S(p^2)$, is illustrated in
Fig.~\ref{fig:figB}. Fig.~\ref{fig:figG} shows the photon
renormalization function, $\G(q^2)$, found from their self-consistent
solution and this is compared with its 1-loop approximation. One
observes that at high momenta the self-consistent $\G(q^2)$ follows
the 1-loop result very nicely. For decreasing momenta the effect of
the dynamically generated mass comes into play and the value of
$\G(q^2)$, and hence that of the running coupling, seems to stabilize
for a while, as one could expect. Then, surprisingly, at some lower
momentum there is a sudden fall in $\G(q^2)$, which drops below
the 1-loop value and almost vanishes completely. This is a rather
strange behaviour for the running coupling at low momenta. This
decrease corresponds to $\tilde{\Pi}(q^2)$ of \mref{vacpolrenorm}
becoming large.


To solve the problem numerically Kondo et al. have made supplementary
assumptions about the ultraviolet behaviour of $\S(p^2)$ and
$\G(q^2)$. These arise from the need to handle loop momenta beyond the
ultraviolet (UV) cutoff.  For example, if in \mref{S2} $0 \le p^2, k^2 \le
\Lambda^2$, then the photon momentum $q^2 = p^2 + k^2 -
2pk\cos\theta$ will lie in the interval $0 \le q^2 \le
4\Lambda^2$. The same argument holds for the fermion momentum $p^2$ in
\mref{G2}, i.e. $0 \le p^2 \le 4\Lambda^2$. As a consequence the
angular integrals need values of $\S$ and $\G$ at momenta above the
UV-cutoff, this is outside the {\it physical momentum region}. Therefore
one will have to extrapolate $\S$ and $\G$ outside this region. In
their work, Kondo et al. define~:
\ba
\Sigma(q^2 > \Lambda^2) &\equiv& 0 \mlab{extrapS} \\
\Pi(q^2 > \Lambda^2) &\equiv& 0  \; \Rightarrow \;  \G(q^2 > \Lambda^2)
\equiv  1 \mlab{extrapG} \, .
\ea

Both dynamical mass and vacuum polarization vanish above the UV-cutoff
and the theory then behaves as a free theory.  Although this
assumption seems reasonable, \mref{extrapS} introduces a jump
discontinuity in the dynamical mass function at $q^2 = \Lambda^2$
because $\S(\Lambda^2) \ne 0$ for $\alpha > \alpha_c$ (see
Fig.~\ref{fig:figB}), while \mref{extrapG} introduces a relatively
sharp kink in the photon renormalization function at that point (see
Fig.~\ref{fig:figG}).

In the physical world these functions have to be smooth. To
investigate in a crude way the influence of the discontinuity in
$\S(p^2)$, we can remove it by hand by defining the following simple
extrapolation rule~:
\be
\S(p^2 > \Lambda^2) = \S(\Lambda^2) \, \frac{\Lambda^2}{p^2} \, .
\mlab{extrapS2}
\ee

This will get rid of the jump discontinuity in the dynamical mass
function, leaving instead a very slight kink.

\begin{figure}
\begin{center}
{}~\epsfig{file=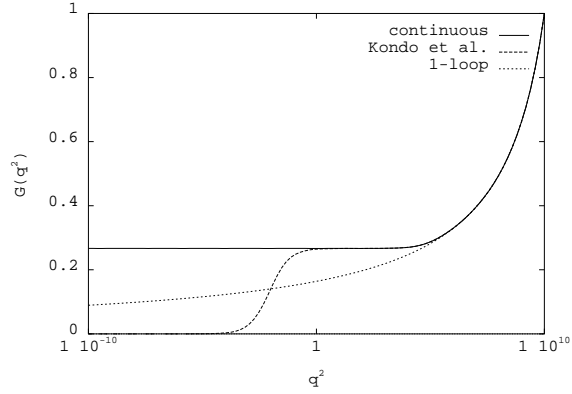,height=8cm,angle=-90}
\end{center}
\vspace{-0.5cm}
\caption{Photon renormalization function $\G(q^2)$, as a function of
momentum $q^2$ for $N_f=1$ and $\alpha=2.086$ as calculated in a
self-consistent way with a continuous extrapolation for $\S(p^2)$,
with the jump discontinuity in $\S(p^2)$ as in \protect{\cite{Kondo92}}
and in 1-loop approximation ($\Lambda = 10^5$).}
\label{fig:figG2}
\end{figure}

When solving the integral equations using this extrapolation rule, the
step in the photon renormalization function at intermediate low
momenta surprisingly disappears as can be seen in
Fig.~\ref{fig:figG2}. This was not anticipated since one would not
expect the high momentum behaviour of $\Sigma(p^2)$, where its value
is quite small, to play such a major role in the behaviour of
$\G(q^2)$ at low $q^2$.

A more detailed investigation indeed shows that the step in the photon
renormalization function found by Kondo et al. is an artefact of the
way they renormalize the quadratic divergence in the vacuum
polarization integral, \mref{G2}, combined with the presence of the
jump discontinuity in the dynamical mass function, \mref{extrapS}, as
we now explain.

 From the angular integrand of the $\G$-equation, \mref{G2}~, we define
$f_\theta$ as~:
\be
f_\theta =
\frac{k^2 - kq\cos\theta + 2\S(k^2)\S(p^2)}{p^2+\S^2(p^2)} -
\frac{k^2+2\S^2(k^2)}{k^2+\S^2(k^2)} \, .
\mlab{angint}
\ee

Both terms in \mref{angint} cancel exactly at $q^2 = 0$ to remove the
quadratic singularity. It is easy to see that provided $\S(k^2)$ is
continuous for all $k^2$, $f_\theta$ will be continuous, and if
$\S(k^2)$ has a Taylor series, $f_\theta$ will be smooth. Of course
the description of the real world has to be such that the approximate
cancellation of the quadratic divergence at low $q^2$ becomes exact at
$q^2\ =\ 0$ in a smooth way.

Now let us look at the angular integrand $f_\theta$ in the
approximation of Kondo et al. \cite{Kondo92} when $q^2$ is small but
$k^2$ is very large, indeed larger than $k_0^2 = (\Lambda - q)^2$. For
values of $\theta$ greater than $\theta_0(k^2) =
\arccos((k^2 + q^2 - \Lambda^2)/2kq)$ we will have
$p^2 > \Lambda^2$. If we now use Kondo et al's extrapolation,
\mref{extrapS}, then $\S(p^2 > \Lambda^2) = 0$ and the angular
integrand \mref{angint}, now becomes~:
\be
f_\theta =
\frac{k^2 - kq\cos\theta}{p^2} -
\frac{k^2+2\S^2(k^2)}{k^2+\S^2(k^2)} \, .
\ee

When $q \rightarrow 0$, i.e. $p \rightarrow k$~:
\ba
f_\theta &\approx& -\frac{\S^2(k^2)}{k^2 + \S^2(k^2)} +
{\cal O}(q^2,kq\cos\theta) \, .
\mlab{ftheta2}
\ea

As soon as $q^2$ deviates from zero, the angular integrand contains a
jump discontinuity at $\theta = \theta_0(k^2)$, and part of the
angular integrand will not vanish continuously when $q^2 \rightarrow
0$. In fact the angular integral $I_\theta$ will receive an extra
contribution $\delta I_\theta$ when $k^2$ is larger than $k_0^2 =
(\Lambda - q)^2$~:
\ba
\delta I_\theta(k^2) &=& \int_{\theta_0(k^2)}^{\pi} d\theta \,
\sin^2\theta \left[ -\frac{\S^2(k^2)}{k^2 + \S^2(k^2)}\right]
\nonumber \\
&=& -\left(\frac{\pi}{2} - \frac{\theta_0(k^2)}{2} + \frac{\sin
2\theta_0(k^2)}{4}\right) \, \frac{\S^2(k^2)}{k^2 + \S^2(k^2)}
\, .
\mlab{dItheta}
\ea

\samepage{
Substituting \mref{dItheta} in \mref{G2} we see that the vacuum
polarization receives an extra contribution $\delta\Pi(q^2)$~:
\be
\delta\Pi(q^2) =
\frac{4N_f\alpha}{3\pi^2}\frac{1}{q^2}\int_{k_0^2}^{\Lambda^2} dk^2 \,
\frac{k^2\S^2(k^2)}{(k^2+\S^2(k^2))^2}\left(\frac{\pi}{2} -
\frac{\theta_0(k^2)}{2} + \frac{\sin2\theta_0(k^2)}{4}\right) \, .
\mlab{extracon}
\ee}

Writing $k=\Lambda + q\cos{\psi}$, so that $\theta_0 \simeq \psi$ for
$q^2 << \Lambda^2$, we have, using the mean value theorem~:
\be
\delta\Pi(q^2) \simeq
\frac{8N_f\alpha}{3\pi^2}\frac{\Lambda^3\S^2(\Lambda^2)}
{q(\Lambda^2+\S^2(\Lambda^2))^2}
\int_{\pi/2}^{\pi} \, d\psi \sin{\psi} \left[\frac{\pi}{2} - \frac{\psi}{2}
+ \frac{\sin2\psi}{4}\right] \, ,
\ee

so that~:
\be
\delta\Pi(q^2) \simeq
\frac{8N_f\alpha}{9\pi^2}\frac{\S^2(\Lambda^2)}{q\Lambda} \, .
\mlab{deltavp}
\ee

Because of the $1/q$ this change in $\Pi(q^2)$ would be noticeable at
very small values of $q^2$. However, this analytic calculation does
not explain the sharp decrease of $\G(q^2)$ at intermediate low
momenta we and Kondo et al. \cite{Kondo92} find --- see
Fig.~\ref{fig:figG}.

To understand why this happens we have to consider how the numerical
program computes the extra contribution \mref{extracon} to the vacuum
polarization integral. The integrals are approximated by a finite sum
of integrand values at momenta uniformly spread on a logarithmic
scale. For small $q^2$ the extra contribution is entirely concentrated
at the uppermost momentum region of the radial integral ($k^2 \in
[k_0^2,\Lambda^2]$). There the numerical integration program will have
only one grid point situated in the interval $[k_0^2, \Lambda^2]$ for
any realistic grid distribution. This point will lie at
$k^2=\Lambda^2$ if we use a closed quadrature formula. Therefore the
integral will be approximated by the value of the integrand at
$\Lambda^2$ times a weight factor $W(\Lambda^2) = w\Lambda^2$ ($w$ is
${\cal O}(1)$)~:
\be
\delta\Pi(q^2) \approx
\frac{4N_f\alpha}{3\pi^2}
\frac{W(\Lambda^2)\Lambda^2\S^2(\Lambda^2)}
{q^2(\Lambda^2+\S^2(\Lambda^2))^2}\left(\frac{\pi}{2}
- \frac{\theta_0(\Lambda^2)}{2} +
\frac{\sin2\theta_0(\Lambda^2)}{4}\right) \, .
\ee

For small $q^2$ we have $\theta_0(\Lambda^2)\approx\pi/2$ and the
extra contribution to the vacuum polarization will be~:
\be
\delta\Pi(q^2) \approx \frac{N_f\alpha
w}{3\pi}\frac{\S^2(\Lambda^2)}{q^2} \, .
\mlab{deltavpnum}
\ee

This will effectively add a huge correction to the vacuum polarization
at low $q^2$. This has been extensively checked numerically and shown
to be completely responsible for the sudden decrease in the photon
renormalization function $\G(q^2)$ at low momenta. To reproduce our
previous analytic result of \mref{deltavp} numerically, the
integration grid has to be tuned unnaturally fine to include more
points in the region $[k_0^2,\Lambda^2]$. Without such tuning one has
the result of \mref{deltavpnum}. Then $q^2\Pi(q^2)$ does not vanish
smoothly as $q^2 \rightarrow 0$. Instead, for $q^2 > 0$, $q^2\Pi(q^2)
\approx N_f\alpha w \S^2(\Lambda^2)/3\pi$ and so as soon as $q^2$ is
non-zero the cancellation of the quadratic divergence disappears
suddenly and not gradually as the physical world requires.

\newpage

How can we avoid this problem? As discussed before one can introduce a
smooth decrease of $\S(p^2)$ above the UV-cutoff. This ensures that
the cancellation of the quadratic divergence takes place smoothly as
$q^2 \rightarrow 0$.  The results obtained with the approximation of
\mref{extrapS2} are shown in Fig.~\ref{fig:figG2} and are consistent
with our physical intuition about the behaviour of the running of the
coupling.

Once the quadratic divergence has been removed properly, other
numerical difficulties start to show up. For instance, inadequate
interpolation may give rise to unphysical singularities in
$\G(q^2)$. We do not discuss these further, as they are outside the
scope of this note. However, we remark that these problems are avoided
if one uses some smooth solution method.

We conclude that one has to ensure the proper removal of the quadratic
divergence from the vacuum polarization integral when solving the
coupled set of integral equations for the dynamical mass function and
the photon renormalization function numerically. As shown, a very
small jump discontinuity in the extrapolation of the dynamical mass
function can alter the behaviour of the photon renormalization
function quite dramatically at low momentum and such a peculiar
running of the coupling is unphysical. To avoid this and also other
numerical problems encountered in the solution of the coupled set of
integral equations it would therefore be preferable to search for
smooth solutions for the dynamical mass function $\S(p^2)$, the
fermion wavefunction renormalization $\F(p^2)$ and the photon
renormalization function $\G(q^2)$. A study implementing this is
currently in progress. This is essential if we are to understand the
phase structure of strong coupling QED in 4 dimensions in the
continuum.

\end{document}